\documentclass[aps,prd,tightenlines,superscriptaddress,nofootinbib]{revtex4}
\usepackage{epsfig}
\usepackage{graphicx}
\usepackage{psfrag}
\usepackage{amsmath,amssymb}
\usepackage{colordvi}
\usepackage{amsfonts}
\usepackage{enumerate}
\usepackage{slashed}
\usepackage{color}
\usepackage{xcolor}

\begin{document}
\title{Reply to arXiv:1211.3957 and arXiv:1211.4731 by Leader {\it et al.} \\ and arXiv:1212.0761 by Harindranath {\it et al.}}
\author{Xiangdong Ji}
\affiliation{INPAC, Department of Physics and Shanghai Key Lab for Particle Physics and Cosmology, 
 Shanghai Jiao Tong University, Shanghai, 200240, P. R. China}
\affiliation{Center for High-Energy Physics, Peking University, Beijing, 100080, P. R. China}
\affiliation{Maryland Center for Fundamental Physics, University of Maryland, College Park, Maryland 20742, USA}
\author{Xiaonu Xiong}
\affiliation{Center for High-Energy Physics, Peking University, Beijing, 100080, P. R. China}
\author{Feng Yuan}
\affiliation{Nuclear Science Division, Lawrence Berkeley
National Laboratory, Berkeley, CA 94720, USA}
\date{\today}
\vspace{0.5in}
\begin{abstract}
We reply to the recent comments on our published papers, Phys. Rev. Lett. 109 (2012) 152005 and Phys. Lett. B717 (2012) 214.
We point out that the criticisms about the transverse polarization parton sum rule we obtained are invalid.
\end{abstract}

\maketitle

The comments by Leader {\it et al.}~\cite{Leader:2012md} and by Harindranath {\it et al.}~\cite{Harindranath:2012wn} 
on our Phys. Rev. Lett. paper~\cite{Ji:2012sj}
arise from a mis-understanding of our result. We have in fact published a longer
paper~\cite{Ji:2012vj} following the Letter, fully explaining what our partonic transverse spin sum rule mean.
We reiterate that our result stands following a careful consideration of all pertinent issues.

First of all, we remarked the simple fact that the ordinary transverse angular momentum (AM)
does not commute with the longitudinal boost, and thus a frame-independent picture for the transverse spin
is not the transverse AM alone, but the well-known Pauli-Lubanski (PL) spin $\hat W_\perp$~\cite{Ji:2012vj}. The PL spin
is diagonalized in the transversely-polarized nucleon state with arbitrary longitudinal momentum.

The PL spin is defined as $\hat W^\mu \sim  \epsilon^{\mu\alpha\beta\gamma} \hat J_{\alpha\beta} \hat P_\gamma$,
and we take the nucleon state with $P^\mu=(P^0,P_\perp=0,P^3)$ and replace
$\hat P^\mu$ by its eigenvalue, so that $\hat W^\mu$ linearly depends on the angular momentum operator
$\hat J^{ij}$, as well as the boost operator $\hat J^{0i}$. In the Letter paper, we restrict
ourselves to the light-cone rest frame with residual momentum $P^3=0$, and thus only $P^+$ and $P^-$ do not vanish.
If taking $\mu=1$, and $\alpha=2$, $\beta, \gamma$ to be $+$ and $-$,
we have $W^1 \sim  -\hat J^{2+}P^- + \hat J^{2-}P^+$. In light-cone quantization,
the $\hat J^{2-}$ is a higher-twist contribution depending on products of three or four parton operators;
however, its matrix element is related to that of $J^{2+}$
by simple Lorentz symmetry, and hence its contribution is considered known. Thus the leading-twist parton picture arises 
from $J^{2+}$ which is interaction-independent. One can obtain a simple partonic interpretation for this part related to the tensor $T^{++}$, 
as explained in the Letter paper, and the result is an integral over the intuitive parton 
transverse AM density $x(q(x)+E(x))/2$  and consistent with Burkardt wave-packet picture~\cite{burkardt}. 
Thus, the key aspect of finding a partonic picture for transverse PL spin is to focus on the leading twist part
and do away the other parts through Lorentz symmetry, a strategy first pointed out by Burkardt. Note that 
the spin operators of quarks and gluons do not contribute at the leading twist 
as they are now higher-twist operators in light-cone quantization.

The PL vector was also the starting point of
Ref.~\cite{Harindranath:2012wn} and an earlier publication of the same
authors, Ref.~\cite{Harindranath:2001rc},
in which the equations (2.6) and (2.7) reduce to $W^i$ when the external particle has
no transverse momentum, $P^i=0$. One can easily find that they
agree with our starting point of the discussion, contrary to the claim in their comment, Ref.~\cite{Harindranath:2012wn}.
Moreover, our conclusion does not contradict with that in Ref.~\cite{Harindranath:2001rc}:
In our longer version of Ref.~\cite{Ji:2012vj}, we find twist-3 and
twist-4 parts of $W^i$ are interacting-dependent.
Our new result~\cite{Ji:2012sj} beyond Ref.~\cite{Harindranath:2001rc} is that
there is a twsit-two contribution of the transverse polarization which can be
understood in a simple parton picture, related to the generalized parton
distributions (GPD), whereas the interaction-dependent part is related
to that of the twist-2 GPD contribution by symmetry.

Finally, Leader in a separate note \cite{Leader:2012ar} criticized our light-front result
when generalized to an arbitrary residual momentum frame~\cite{Ji:2012vj}. A careful reading
of our paper reveals that we have already commented on the role of higher term $\bar C$ in the paragraph
following Eqs. (16) and (23). The frame-independence of our result remains to be true for
the leading twist part, which is a consequence of the dependence of the transverse spin $\hat W_\perp$
on the very boost operator that serves to cancel the frame dependence of the transverse AM.


\begin{thebibliography}{99}


\bibitem{Leader:2012md}
  E.~Leader and C.~Lorce,
  arXiv:1211.4731 [hep-ph].

\bibitem{Harindranath:2012wn}
  A.~Harindranath, R.~Kundu, A.~Mukherjee and R.~Ratabole,
  arXiv:1212.0761 [hep-ph].

\bibitem{Ji:2012sj}
  X.~Ji, X.~Xiong and F.~Yuan,
  Phys.\ Rev.\ Lett.\  {\bf 109}, 152005 (2012)
  [arXiv:1202.2843 [hep-ph]].

\bibitem{Ji:2012vj}
  X.~Ji, X.~Xiong and F.~Yuan,
  Phys.\ Lett.\ B {\bf 717}, 214 (2012)
  [arXiv:1209.3246 [hep-ph]].

\bibitem{burkardt}
  M. Burkardt, Phys.Rev. D72, 094020 (2005) [hep-
ph/0505189].


\bibitem{Harindranath:2001rc}
  A.~Harindranath, A.~Mukherjee and R.~Ratabole,
  Phys.\ Rev.\ D {\bf 63}, 045006 (2001).


\bibitem{Leader:2012ar}
  E.~Leader,
  arXiv:1211.3957 [hep-ph].


\end{thebibliography}
\end{document}